%% file: paper.tex
\makeatletter\def\input@path{{latex/styles/}}\makeatother % Set load paths
\PassOptionsToPackage{table}{xcolor}

\documentclass[letterpaper,twocolumn,10pt]{article}
\usepackage{usenix2024_SOUPS}

\usepackage{fast-todo}
\usepackage{formatting}

\usepackage{adjustbox}
\usepackage{multirow}
\usepackage{booktabs}

\newcommand*\rot{\rotatebox{90}}
\begin{document}
%-------------------------------------------------------------------------------

%don't want date printed
\date{}

% make title bold and 14 pt font (Latex default is non-bold, 16 pt)
\title{A Large-Scale Survey of Password Entry Practices on Non-Desktop Devices}

\def\plainauthor{Anonymous Author}

\author{
{\rm John Sadik}\\
The University of Tennessee, Knoxville
\and
{\rm Scott Ruoti}\\
The University of Tennessee, Knoxville
} % end author

\maketitle
\thecopyright

\input{chapters/abstract}
\input{chapters/intro} 
\input{chapters/rw}

\input{chapters/method}
\input{chapters/quant}
\input{chapters/qual}
\input{chapters/discussion}
\input{chapters/conclusion}

\bibliographystyle{plain}
\bibliography{latex/bibtex/publications,latex/bibtex/authentication,paper}

\appendix

\input{chapters/appendix-survey}
\input{chapters/appendix-demographics}
\input{chapters/appendix-statistics}
\input{chapters/appendix-passwords}

\end{document}

%% file: chapters/abstract.tex
\begin{abstract}
	Password managers encourage users to generate passwords to improve their security.
	However, research has shown that users avoid generating passwords, often giving the rationale that it is difficult to enter generated passwords on devices without a password manager.
	In this paper, we conduct a survey ($n=999$) of individuals from the US, UK, and Europe, exploring the range of devices on which they enter passwords and the challenges associated with password entry on those devices.
	We find that password entry on devices without password managers is a common occurrence and comes with significant usability challenges.
	These usability challenges lead users to weaken their passwords to increase the ease of entry.
	We conclude this paper with a discussion of how future research could address these challenges and encourage users to adopt generated passwords.
\end{abstract}

%% file: chapters/intro.tex
\section{Introduction} \label{ch:intro}

Even with their many issues, passwords remain the dominant form of authentication~\cite{florencio2007large, dellamico2010password, das2014tangled, pearman2017lets, verizon2021data, stobert2018password}.
%Many of these challenges arise due to the difficulty of creating strong passwords that are hard for an adversary to guess but also easy for the user to remember~\cite{riley2006password,dellamico2010password}.
%If a user does succeed in creating such a password, they are likely to reuse that password across multiple websites~\cite{securityscorecard, verizon2021data, wash2016understanding, gaw2006password, notoatmodjo2009passwords}.
%
Password managers attempt to improve users' password hygiene by assisting them with the password lifecycle~\cite{oesch2020that}: generating, storing, and auto-filling passwords.
If used appropriately, password managers can significantly strengthen users' security.

Research has consistently shown that users appreciate the usability benefits provided by password managers~\cite{fagan2017investigation,pearman2019why,ray2021why}.
However, there is growing evidence that shows that users are unwilling to use generated passwords~\cite{pearman2019why}.
For example, one study found that only one in four password manager users use a password generator~\cite{lyastani2018better}.
One reason users give for avoiding generated passwords is that they are difficult to enter on devices where the password manager is unavailable~\cite{oesch2022observational}.

To help solve this problem, password generation algorithms could take into account the devices where passwords will be entered, tailoring the generated password to be easy to enter on those devices, even if password autofill is unavailable.
To build such algorithms, it is necessary to first answer the following two research questions:

\begin{description}
	\item[RQ1] On which devices do users enter passwords and how frequently do they do so?
	\item[RQ2] What challenges do users face when entering passwords on non-desktop devices?
\end{description}

In our search of the literature, we found that RQ1 remains entirely unanswered and RQ2 has only been partially answered for touchscreen devices (i.e., tablets and smartphones).
Critically, this knowledge gap prevents the principled design of device-aware password generators (our ultimate research goal).
As such, in this work, we seek to fill this knowledge gap by surveying 999 participants, gathered from the US, UK, and Europe, asking them about (i) what devices they authenticate on, (ii) the frequency of that authentication, and (iii) the challenges they face authenticating on those devices.
Key findings from our research include,

\begin{enumerate}
	
	\item
	Our results show that users do need to authenticate on a variety of devices, many of which do not support the use of a password manager to autofill passwords.
	These findings highlight the need for research that examines the usability of authentication across a range of less common devices---not just in the case of passwords, but for all authentication schemes (e.g., hardware security keys).
	
	\item
	We also find that the input characteristics of the devices used to enter passwords have a significant impact on the usability of password entry.
	Critically, when password entry is difficult on a given device, most participants report simplifying (i.e., weakening) the passwords they expect to enter on that device.
	These findings help explain why users would choose to reject generated passwords, as generated passwords would be unwieldy to enter on many devices.
	This highlights the importance of creating device-aware password-generation algorithms.
	%	In response to this finding, we also contribute some leads on future work that can address this issue. 
	
	%This includes the (intuitive) result that some devices are faster or easier to enter credentials on.
	%More importantly, we demonstrate that these differences in usability directly impact how users create passwords, with over half of our participants admitting to changing their passwords based on the device where they will enter the password.
	%Our qualitative data shows that these changes uniformly lead to weaker passwords.
	
	% JOHN: I took this out because I feel like reviewers will complain about it.
	%\item
	%We explore users' frustration with passwords.
	%For example, a quarter of participants mentioned the challenge of remembering passwords is the greatest challenge they face with authentication.
	%Similarly, participants indicated that their frustrations are heightened by the large number of websites on which they need to create passwords and the existence of burdensome password composition policies.
	%Many of the challenges we identify can be easily fixed by websites and frameworks.
	%Based on our other findings, we believe this increase in usability could have a commiserate increase in the security of passwords that users select.
	
	\item
	Lastly, we find that our participants are very interested in adopting and using password managers and other tools for authentication, such as biometrics.
	%In fact, one in seven of our participants already uses a password manager.
	However, we also identify problems with users' understanding of these technologies, causing a small, but non-negligible fraction of participants to distrust password managers and other authentication tools.
	As such, there is a need to better educate users about how these technologies work, especially relating to how their data is handled, in an effort to increase adoption. %, though doing so is no small task.
	
\end{enumerate}

We conclude this paper by discussing the implications of our results on building authentication systems, and in particular building device-aware password generation algorithms.

%Although some of our findings are intuitive, they are still important as they provide empirical backing for these findings.

%% file: chapters/rw.tex
\section{Related Works}
\label{ch:rw}

In reviewing the literature, we did not find any research specifying the range of devices that users use to authenticate, nor the frequency at which they did so (\textbf{RQ1}).
Additionally, while there is some research investigating challenges with password usage on touchscreens, there was still a knowledge gap involving challenges for password entry on non-desktop, non-touchscreen devices (\textbf{RQ2}).
This knowledge gap is the motivating factor behind this work.

In this section, we start by describing research into password manager usage, research that motivates the need for device-aware password generation.
We then describe what research there is regarding textual input on touchscreens and TVs, including the work by Greene et al.~\cite{greene2014can,greene2015tap} and Jakobsson et al.~\cite{jakobsson2012rethinking} exploring challenges with password entry on touchscreen devices.

\subsection{Password Manager Usage}

Research has shown that usability, not security, is often the primary motivation for users to adopt a password manager~\cite{fagan2017investigation,pearman2019why,ray2021why,alodhyani2020password}.
In fact, security concerns can lead users to eschew adopting a password manager~\cite{fagan2017investigation,ray2021why,alodhyani2020password}.
\emph{Based on our findings in this paper, we believe that some of this fear regarding the security of password managers arises from users' poor mental models regarding password managers.}

This focus on usability---not security---likely helps explain why even when users adopt a password manager, they often ignore security-critical functionality such as password generation~\cite{lyastani2018better,oesch2022observational}.
When asking directly about this issue, Oesch et al.~\cite{oesch2022observational} found that users reported avoiding generated passwords due to concerns related to remembering and entering these passwords on devices where the password manager was not available, especially those devices without physical keyboards.
\emph{Our findings show that users commonly enter passwords on these types of devices, providing support for the idea that password managers should consider the entry device when deciding how to generate a password.}

\subsection{Touchscreen Entry}

%There is a substantial body of research on textual entry on touchscreens.
%Here we describe the items most relevant to our study.

Research into textual entry on touchscreens shows that it is slightly slower than using a physical keyboard---roughly 15--30 words per minute using a touchscreen~\cite{karat1999patterns,mackenzie2002mobile} as opposed to 40 words per minute using a keyboard~\cite{type,ostrach1997typing}.
Errors when using a touchscreen keyboard are also high~\cite{lee2009soft}.
Textual entry on a mobile device using voice transcript is even slower, at roughly 14 words per minute, largely owing to the higher rates of mistakes and the need to correct those mistakes~\cite{karat1999patterns}.

When it comes to manual password generation on touchscreens, Yang et al.~\cite{yang2014text} show that users were more likely to use lowercase letters in passwords for their smartphones. 
Melicher et al.~\cite{melicher2016usability} show that these user-created passwords on mobile devices are weaker against strong attackers, but against a less strong attacker, the difference is not discernible. 
\emph{Importantly, in our study we find that users are admitting that they are weakening their passwords. 
	Even if the research shows these passwords are not actually much weaker, it is interesting that users think they are weaker and are still consciously choosing to make this trade-off.}

Looking more specifically at password entry, Von Zezschwitz et al.~\cite{von2014honey} found that entry time was long, up to 22 seconds.
As expected, the entry time increases as the complexity of the input increases.
Additionally, users frequently make errors when entering passwords on mobile devices, which not only helps explain the higher entry time but could induce frustration.

Greene et al.~\cite{greene2015tap} achieved positive preliminary results when they attempted to improve password entry on touchscreens by investigating whether password generation could be modified to group characters from the same character class, reducing the number of times users needed to switch the virtual keyboard's layout.
Jakobsson et al.~\cite{jakobsson2012rethinking} also attempted to improve password entry on touchscreens through the use of passphrases.
As passphrases are composed of dictionary words, they benefit from the error correction found on many touchscreen keyboards.
In a user study evaluating this approach, Jackobsson et al. found that the passphrases they studied were quicker to enter than standard passwords and had a higher recall rate.
\emph{Our research shows that many users need to semi-frequently enter passwords on touchscreen devices, suggesting that there is a need for additional research along the lines of Greene et al. and Jakobsson et al. to tailor generated passwords for usage on touchscreens.}

%There has also been other research~\cite{khamis2016gazetouchpass,bianchi2010phone,bianchi2011spinlock,bianchi2012counting,liu2015exploiting} which provide alternatives to password entry with multimodal schemes of authentication. 
%The highlight here is that these schemes are usually more secure than some existing methods while also being usable, even though they incorporate new input modalities, which is a goal this area of research strives for. 

\subsection{TV Entry}
Bobeth et al.~\cite{bobeth2014tablet} conduct a study that compares standard remote controller entry, gesture-based entry (which was a wizard-of-Oz type entry), and a screen-mirrored tablet for entry as different ways to enter text on a TV. 
In their study, they look at how different age groups use each input modality and what the impact on that age group's user experience is. 
Bobeth et al. found that older users had worse motor skills, and therefore it took longer for them to complete the tasks in the user studies. 
However, the more interesting result is that neither the application used nor the age of the participant had an impact on usability. 
The only thing that had an impact on usability in this study was the input modalities. 
This work also found that avoiding display switching seemed to be advantageous in this context. 

The work of Coelho et al.~\cite{coelho2011developing} and Simon et al.~\cite{simon2013enrichment} explore many different input modalities related to using a TV. 
These input modalities include speech, a separate touchscreen screen for input, finger-pointing, and gestures. 
The goal of this exploration is to explore the usability of different input modalities, especially as it relates to the elderly and children. 
This is because the work of Li et al.~\cite{li2010mems} showed that the standard remote control entry with many buttons, which remains the default entry method~\cite{pirker2010investigating}, is difficult for the elderly and children. 
This is especially important because the work of McLaughlin et al.~\cite{mclaughlin2009using} provides empirical evidence that the input modality used impacts performance.

%% file: chapters/method.tex
\section{Methodology} \label{ch:method}
We conducted a survey of users' authentication experiences on different devices.
This survey was conducted and completed on October 10th, 2022.
In total, we collected 1,003 responses, and after removing four for data quality issues, we were left with 999 valid responses.

We sent this survey out to three different regions: Europe (n=601), the United States (n=229), and the United Kingdom (n=99). 
The difference in participant count is based on the relative populations of each region.

The survey was distributed using Prolific and administered using Qualtrics.
Each participant could only take the survey once and was provided USD \$1.50 for their participation, resulting in an average of \$14.44/hour. 
The survey was approved by our Institutional Review Board and the survey instrument is reproduced in Appendix~\ref{appx:survey}.

\subsection{Survey Content}
We started the survey by briefing participants on the nature of the study and collecting their informed consent.
We also informed users that during the study we would use the following definition for authentication: ``The process of logging into an account is referred to as authentication.''
We asked only about authentication with a password or PIN because we were interested in the impact of the devices on authentication, so we limited the scope appropriately.

In the survey, we first asked participants which devices they have used to authenticate. 
They were able to select from a list of general devices (e.g. laptops), gaming devices (e.g. Xbox), smart devices (e.g. smart TV), and physical devices (e.g. ATMs). 
Users were also able to manually enter other devices they had authenticated on but that were not in the previous questions. 

Second, we asked participants which three to five devices they had used most frequently to authenticate. 
Third, we asked them how often they used different input modalities to authenticate. 
Fourth, we asked participants to reflect on the intersection of the device they were using and the password they were using.
Fifth, we asked participants open-ended questions about any challenges they faced and any additional comments they had. 
Finally, we collected demographic data.

\subsection{Survey Development}

Keeping our research questions in mind, we created our survey and revised it through repeated rounds of iteration.
In each iteration round, we considered our research questions and our current survey and sought to modify our survey to better answer those research questions.
We first focused on the list of devices users used and how often they used those devices. 
This way, as they progressed through the survey, more devices were on the user's mind.
Then we asked users directly about areas of impact we thought of using our research questions. 
Finally, we allowed users to share their comments with us.

Once we produced a version we were satisfied with, we submitted for and received IRB approval for our study.
Next, we conducted a pilot study using a convenience sample of ten participants.
These participants included both technical and non-technical users, and they were simply instructed to take the survey and then share their feedback. 
After considering their feedback and reviewing their answers, we did not detect any significant problems with our survey nor were there any misunderstandings of our survey questions, so we launched the final survey.

Using Prolific, we created three copies of our survey, each targeting a different region: Europe, the United States, and the United Kingdom.
We selected the target participant counts based on the relative size of the populations in each of these areas.
All participants were required to have proficiency in English, though it did not need to be their primary language.
This was required as the survey was in English.
We also used Prolific to balance the gender of respondents to our survey.

Originally, our survey contained additional questions that asked about devices used and the frequency of authentication with biometrics.
Unfortunately, even though these questions were included in the pilot study, an error made by one of the authors caused them to be omitted in the final study.
However, many participants still commented about biometrics in their answers, and those comments are analyzed and reported. 

\subsection{Quality Control}
While reviewing the data, we looked for responses that indicated participants weren't properly engaging with the survey---for example, using the question as the answer, providing an incoherent or off-topic answer, or failing to provide an answer for a majority of the questions. 
We did not use attention checks as, based on prior experience, we felt that it would be clear who was paying attention based on the quality of their open-response questions. 

In the end, only four participants (0.4\% of our data) had such answers for a majority of the open-ended or text-based questions, and after careful review from both coders, these responses were removed from the dataset. 
The remaining 999 responses comprise the results of our survey, and these are the only responses considered in the analysis of the data. 

When analyzing the quality of responses, the coders also considered whether participants had correctly understood that our survey was measuring password entry \textit{on} the devices, as opposed to using a password to log \textit{into} the device.
While there was some ambiguity in a small number of responses ($<$1\%), the coders agreed that based on participants' answers to open ended question, participants overwhelmingly understood what was being asked.

\subsection{Demographics}
We had an even mix of males (50\%) and females (45\%), with participants skewing younger: 18--25 (37\%), 26--35 (34\%), 36--45 (16\%), 46+ (13\%)).
Participants were well educated: High school graduates (38\%), college graduates (39\%), and advanced degree (20\%).
%As for education, 597 (60\%) respondents said that they had completed college with a degree, 972 (97\%) said that they had at least completed high school or some equivalent, and the remaining respondents (27, 27\%) either had not completed high school or its equivalents or they preferred not to respond.
Appendix~\ref{appx:demographics} gives a full breakdown of demographic data.

\subsection{Quantitative Data Analysis}
When analyzing the quantitative data, we used a $\chi^2$ test to look for differences in the results based on where the respondents came from (Europe, USA, UK).
%We did this by using the tabular data by looking at the number of people who reported using each device. 
In most cases, the differences were not statistically significant.
Even when they were, the effect size was so small as to be negligible.
As such, we do not break down our results based on nationalities.

\subsection{Qualitative Data Analysis}
We gathered qualitative data from the three open-ended questions in our survey.
To analyze this data, we used a methodology inspired by grounded theory~\cite{strauss1997grounded}.
All coding was completed by two researchers who were present at all stages of the process.

First, these coders read through each response together, applying open coding to generate an initial set of codes describing the data.
If there were ever disagreements about the codes to assign, the coders would discuss them until they came to an agreement.
For this reason, we do not report any intercoder reliability, because all coding was done with both coders, and codes were only reported after reaching perfect agreement. 
Throughout this process, they used the constant comparative methods~\cite{glaser1965constant} to identify codes that were originally separate and could be combined, as well as codes that were originally combined and should be split.

Second, after open coding, the coders proceeded to axial coding.
In axial coding, similar codes (both within and between questions) are grouped into ``concepts''.
The coders then grouped these concepts into themes, describing how the concepts each related to the theme and each other.
These concepts and themes are reported in \S\ref{ch:qual}.

The coders do not continue through selective coding (the final step of grounded theory).
Throughout this process, the two coders kept a detailed set of research notes.
These notes aided in the process of coding but also included insights and lessons learned as the coders completed the coding process.
According to grounded theory, these notes are often just as valuable as the actual codes.
Many of our findings are contextualized based on the insights found in the research notes.

% NOTE: I'm not sure this is really needed, but I left it in since you (John) added it.
\paragraph{Input Modalities and Interfaces}
Previous work has shown that there is a high degree of correlation between input modalities and interfaces.
For example, screen size directly impacts the virtual keyboard interfaces shown to users~\cite{greene2014can,greene2015tap,von2014honey}.
Similarly, the need to use arrow keys to navigate a virtual keyboard impacts the design of the virtual keyboard~\cite{bobeth2014tablet,li2010mems,pirker2010investigating}.
Because these two concepts are so closely intertwined, we analyzed them together as a single concept.

%% file: chapters/quant.tex
\section{Quantitative Results}\label{ch:quan}
We first report our findings about what devices are used to authenticate, the frequency of that authentication, and the impact devices have on usability.
These results are based on the closed-ended questions from our survey and work together to answer \textbf{RQ1}.\footnote{RQ1: On which devices do users enter passwords and how frequently do they do so?}

\subsection{Device Usage}

\begin{table}[t]
	\centering
	\adjustbox{max width=\columnwidth}{
		\begin{tabular}{ll|r@{\hspace{15pt}}lr|}
			\multicolumn{2}{l|}{} & \multicolumn{3}{c|}{\shortstack[c]{Count\\(\color{blue} \% Within \color{black}; \% Overall)}} \\
			
			\midrule
			\multirow[c]{6}{*}{\rot{General}} & Any general device & \multicolumn{3}{c|}{997 (100\%)} \\ \cmidrule{2-5}
			& Phone & 980 & (\color{blue} 98\%\color{black};  & 98\%) \\
			& Laptop & 847 & (\color{blue} 85\%\color{black};  & 85\%) \\
			& Desktop & 644 & (\color{blue} 65\%\color{black};  & 64\%) \\
			& Tablet & 481 & (\color{blue} 48\%\color{black};  & 48\%) \\
			& Smartwatch & 127 & (\color{blue} 13\%\color{black};  & 13\%) \\
			% & Smart speaker & 36 & (\color{blue} 4\%\color{black};  & 4\%) \\
			\midrule
			
			\multirow[c]{5}{*}{\rot{Physical}} & Any physical device & \multicolumn{3}{c|}{862 (86\%)} \\ \cmidrule{2-5}
			& ATM & 819 & (\color{blue} 95\%\color{black};  & 82\%) \\
			& Physical keypad & 379 & (\color{blue} 44\%\color{black};  & 38\%) \\
			& Kiosk computer or tablet & 168 & (\color{blue} 19\%\color{black};  & 17\%) \\
			& Printer & 136 & (\color{blue} 16\%\color{black};  & 14\%) \\ \midrule
			
			\multirow[c]{5}{*}{\rot{Smart}} & Any smart device & \multicolumn{3}{c|}{570 (57\%)} \\ \cmidrule{2-5}
			& TV / Smart TV & 416 & (\color{blue} 73\%\color{black};  & 42\%) \\
			& Security alarm & 165 & (\color{blue} 29\%\color{black};  & 17\%) \\
			& Lock / Smart lock & 141 & (\color{blue} 25\%\color{black};  & 14\%) \\
			& Safe / Smart safe & 112 & (\color{blue} 20\%\color{black};  & 11\%) \\
			%			& Thermostat / Smart thermostat & 34 & (\color{blue} 6\%\color{black};  & 3\%) \\
			\midrule
			
			\multirow[c]{7}{*}{\rot{Gaming}} & Any gaming device & \multicolumn{3}{c|}{447 (45\%)} \\ \cmidrule{2-5}
			& PlayStation & 265 & (\color{blue} 59\%\color{black};  & 27\%) \\
			& Xbox & 180 & (\color{blue} 40\%\color{black};  & 18\%) \\
			& Nintendo Switch & 178 & (\color{blue} 40\%\color{black};  & 18\%) \\
			& VR headset & 36 & (\color{blue} 8\%\color{black};  & 4\%) \\ \midrule
			%			& Other game console & 23 & (\color{blue} 5\%\color{black};  & 2\%) \\
			%			& Steam Deck & 22 & (\color{blue} 5\%\color{black};  & 2\%) \\ \midrule
			
			& Any other device & \multicolumn{3}{c|}{94 (9\%)} \\
			\midrule
		\end{tabular}
	}
	\caption{The list of the most common devices on which users have used to enter a password or PIN. Percentages recorded include the percentage within the category and the percentage of overall responses (\% Within; \% Overall).}
	\label{tab:device_usage}
\end{table}

Table~\ref{tab:device_usage} summarizes the devices users have used to enter a password or PIN.
%As noted in our Limitations (\S\ref{sec:limitations}), almost all users understood that our questions about authentication related to logging into an account and not into their device. 
These results make it clear that there is a wide array of devices used to authenticate.
While many of these device types are to be expected, others are more surprising, such as nearly half of all participants authenticating on smart devices and game consoles.

Another interesting observation is that only four of these devices (phone, laptop, desktop, tablet) allow the installation and use of a password manager to support password autofill.
If users were to generate all their passwords, they would need to manually enter the generated passwords on many devices.
As such, there is a strong need to support this use case, though most modern managers do not~\cite{simmons2021systematization}, helping to explain user hesitancy to use generated passwords~\cite{lyastani2018better,oesch2022observational}.

\subsection{Usage Frequency}

\begin{figure}[t]
	\centering
	\includegraphics[width=\columnwidth]{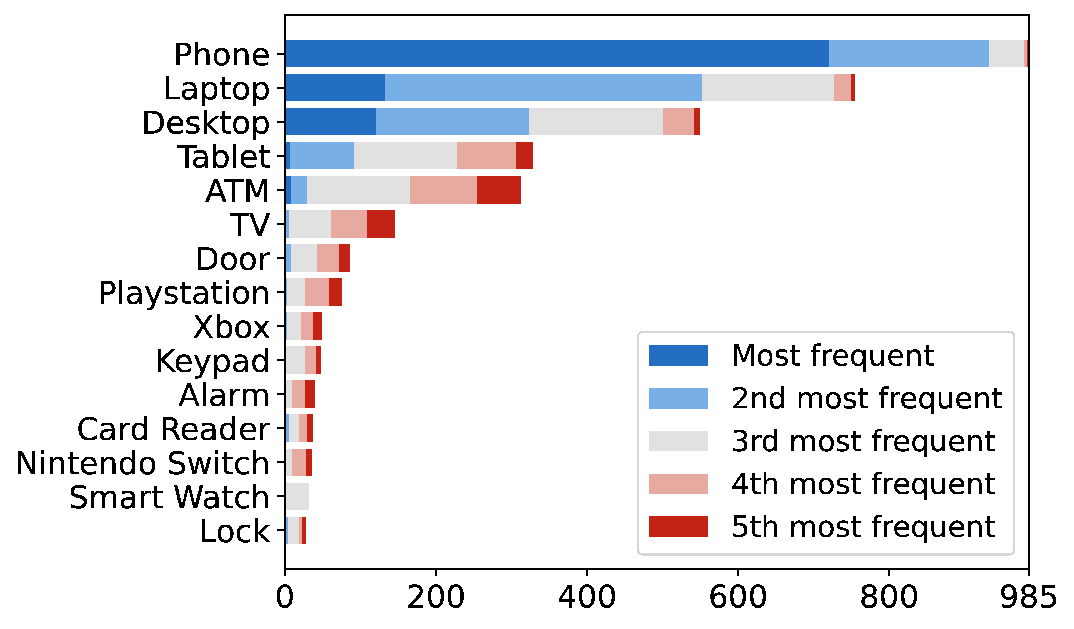}
	\caption{Devices Used for Authentication by Frequency}\label{fig:freq}
\end{figure}

Figure~\ref{fig:freq} lists the relative frequency that users report using their devices to enter passwords or PINs.
As expected, the most common devices are phones, laptops, and desktops.
However, we were surprised to see that phones are the most common place authentication occurs, beating out traditional computing devices (desktops and laptops).
This is surprising when we take into account all the participants' comments complaining about virtual keyboards and all the comments praising desktops and laptops.
We were also surprised by the high usage of ATMs and gaming devices.

\begin{figure}[t]
	\centering
	\includegraphics[width=\columnwidth]{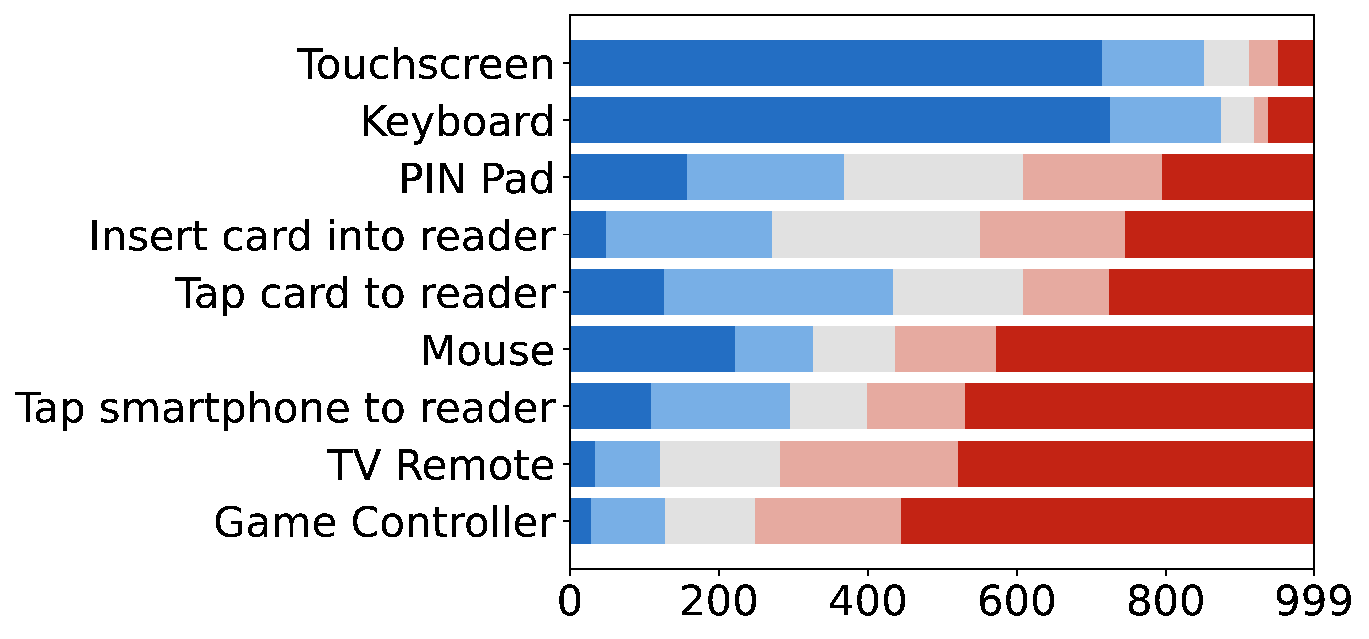}
	\includegraphics[width=\columnwidth]{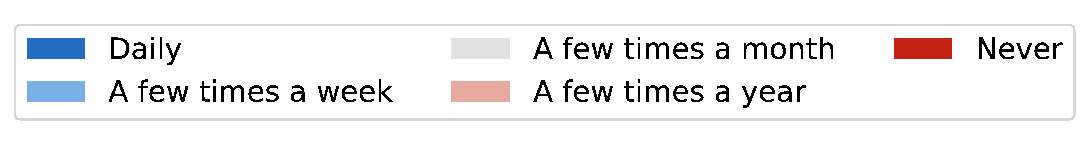}
	\caption{Device Authentication Frequency}
	\label{fig:device_usage}
\end{figure}

Figure~\ref{fig:device_usage} shows how frequently the various input modalities associated with each device are used to enter a password or PIN.
As expected, password or PIN entry occurs very frequently using touchscreens and keyboards.
Due to the prevalence of ATM usage, it is also no surprise that PIN pads are a commonly used input modality.

More surprising is the high number of users that reported using a mouse to enter credentials (n=572). 
We are only aware of a single website (\url{https://www.treasurydirect.gov/}) that requires users to enter a password with their mouse, but there are clearly others.
This may be in reference to the use of on-screen keyboards or captchas that require the use of a mouse to make a selection. 
Future research should certainly examine this phenomenon more to understand in what way users are using a mouse to authenticate.

We were also surprised with the frequency at which users need to enter passwords or PINs using smart TVs or game controllers.
For TV remotes, a quarter of participants (n=282) enter credentials a few times a month, and a half (n=521) do so a few times a year.
The rates are similar for game controllers (n=248 and n=445, respectively).
As we will discuss later, entry on these devices is difficult and frustrating for users.

\subsection{Impact on Usability}

\begin{figure}
	\centering
	\includegraphics[width=\columnwidth]{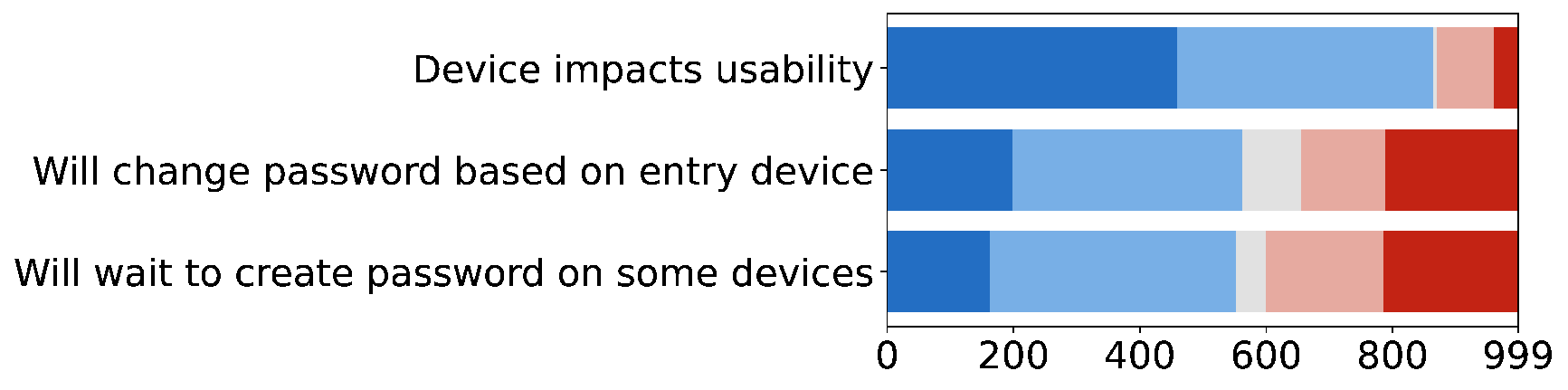}
	\includegraphics[width=\columnwidth]{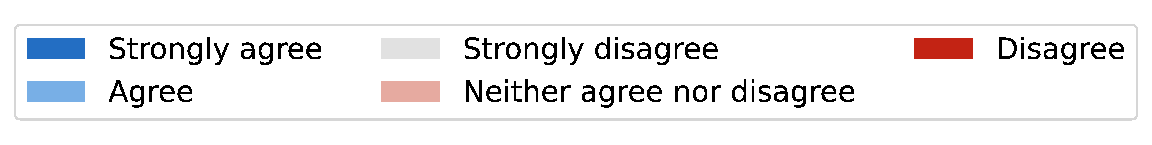}
	\caption{How Much Does Usability Impact User Experience}\label{fig:usability}
\end{figure}

Figure~\ref{fig:usability} indicates how devices impact users' experience with passwords.
As expected, nearly nine-in-ten (n=864) participants feel that password and PIN entry is easier on some devices than on others.
The impacts of this reality are important.
Half (n=562) of participants indicate that they select passwords differently based on what devices will use to enter the password.
As discussed later in this paper, this mostly means that participants weaken their passwords when needing to enter them on non-keyboard devices.
Similarly, half (n=553) of participants will wait to create a password (and the associated account) depending on the device they are using.

%% file: chapters/qual.tex
\section{Qualitative Results}\label{ch:qual}

Next, we discuss the qualitative results of our survey. 
These results help shed light on the challenges users face when entering passwords and authenticating using non-desktop devices, helping answer \textbf{RQ2}.\footnote{RQ2: What challenges do users face when entering passwords on non-desktop devices?}

When interpreting this data, it is important to remember that codes are based on self-elicited responses from participants.
That means that even if only a small fraction of participants complain about a given issue, there are likely many more who feel the same way.
As such, our results identify lower bounds for these issues and should not be interpreted as upper bounds or exact estimates.

\subsection{Devices' Impact on Usability}
\label{sec:usability}

\begin{table}
	\centering
	\adjustbox{max width=\columnwidth}{
		\begin{tabular}{ll|rr@{\hspace{15pt}}l|}
			&& \multicolumn{3}{c|}{\shortstack[c]{Count\\(\color{blue}\% Theme\color{black}; \% Overall)}} \\
			\midrule
			&Devices has some impact & \multicolumn{3}{c|}{855 (86\%)} \\
			&Device has no impact & \multicolumn{3}{c|}{53 (5\%)} \\
			\midrule
			
			\multirow{6}{*}{\rot{Specific impacts}}
			&Any comment & \multicolumn{3}{c|}{336 (34\%)} \\ \cmidrule{2-5}
			&Speed & 143 & (\color{blue}43\%\color{black};  & 14\%) \\
			&Perceived security & 83 & (\color{blue}25\%\color{black};  & 8\%) \\
			&Size & 69 & (\color{blue}21\%\color{black};  & 7\%) \\
			&Mistakes & 57 & (\color{blue}17\%\color{black};  & 6\%) \\
			&Familiarity & 43 & (\color{blue}13\%; \color{black}  & 4\%) \\
			&Comfort & 11 & (\color{blue}3\%; \color{black}  & 1\%) \\
			\midrule
		\end{tabular}
	}
	\caption{Themes and their codes regarding the device's impact on usability and password selection. Percentages for codes are reported based on the percentage within the theme and overall.}
	\label{tab:general_usability}
\end{table}

Table~\ref{tab:general_usability} reports on how participants felt that devices impacted the usability of entering passwords or PINs.
As with any theme reported in this paper, we only coded responses that clearly denoted a stance.
Responses that were ambiguous or did not talk about a device's impact were not coded one way or the other. 
Nearly all participants (n=855) stated that there was some impact, with only 5\% (n=53) indicating that there was no impact.
As we discuss the different ways in which devices impact usability in this section, it is important to keep in mind that in section \ref{ch:quan} we mentioned two of the concrete ways in which this impact might be realized: changing the password composition and delaying the use of a service or device. 

\paragraph{Speed}

The most common impact of the device was the speed at which the password could be entered, with one-seventh of participants (n=143) mentioning this topic.
Notably, participants were most likely to say that entry on a keyboard was fast (n=47), with none saying this entry method was slow.

When discussing the speed of entries, participants voiced strong opinions.
In several cases, participants indicated that if entering passwords was slow, they would avoid using services that required password entry on those devices:

\begin{quote}
	\textit{``the longer it takes, the more annoying it is and i use it less''} (R23)
\end{quote}

\begin{quote}
	\textit{``according to the device I am using, it is more or less quick and easy to authenticate. when it's complicated it's frustrating and sometimes I just give up''} (R387)
\end{quote}  

\paragraph{Mistakes and Device Size}
Mistakes were another common (n=57) frustration.
Often these mistakes were caused by the device being too small, another common concern (n=69):

\begin{quote}
	\textit{``I find it hard to use the touchscreen keyboard on a phone as it is so small. I sometimes get my passwords wrong because of accidentally pressing the wrong button. I would like it to be easier, but I don't know how it could be made easier.''} (R320)
\end{quote}

The other impact of device size is that it can make it difficult to read and interact with the authentication interface:

\begin{quote}
	\textit{``I don't like to authenticate anything on my phone as I can't see everything properly-[w]orried to make a mistake.''} (R75)
\end{quote} 

\begin{quote}
	\textit{``It is preffered to use a laptop to enter passwords as it is easier to observe pop ups or other unwanted elements on the screen in comparison to phone or other similar device[s]''} (R77)
\end{quote} 

\paragraph{Security Concerns}
Finally, a tenth of participants (n=83) indicated that they had concerns regarding using certain devices to authenticate.
While most of these security concerns were left vague, shoulder surfing was mentioned repeatedly (n=14).
Concerns about shoulder surfing extended to both mobile device usage and entering passwords onto large screens:

%\begin{quote}
%	\textit{``I am also afraid that someone might be trying to see the password.''} (R184)
%\end{quote} 

\begin{quote}
	\textit{``So for example.. I can enter my phone pin very quickly, and only I'm able to view the screen generally. If I'm entering my password on say.. Xbox Live, and other people are in the room - they could, if they wanted to, just watch which keys I was hitting on the onscreen keyboard. So because of this my Xbox password is shorter so that I can enter it as quickly as possible''} (R35)
\end{quote} 

\begin{quote}
	\textit{``Entering passwords on some devices (those that are displayed via the TV) are not as safe as those on a computer or smartphone as everyone can see the keys that are being entered. This isn't very secure.''} (R151)
\end{quote}

\paragraph{Statistics}
We used a $\chi^2$ test to investigate whether the difference in feedback between devices was statistically significant, finding that they were ($\chi^2(28,8)=641.6; p<0.001$).
Using a series of pairwise $\chi^2$ tests, we find that the difference between all pairs is statistically significant (see Table~\ref{tab:chi-squared} in Appendix~\ref{appx:chi}),\footnote{$\alpha=0.005$ after applying the appropriate Bonferroni correction.} except between (i) gaming devices and TVs, and (ii) touchscreen and mobile devices.
One reason game devices and TV remotes might have had largely the same comments is that they share an input modality: arrow keys. 
Similarly, touchscreen and mobile devices share an input modality as well. 

\subsection{Devices' Impact on Password Composition}

A device's impact is not solely limited to user experience but also impacts security.
A small but significant number of participants (n=60) indicated that they modify their passwords based on the devices where they will use them (see Table~\ref{tab:general_usability}).
In each case, the implication was that they chose weaker passwords when it was hard to enter the password:

\begin{quote}
	\textit{``If I'm creating an account in a device like a TV, where entering a password takes too long, I might make it shorter or simpler''} (R459)
\end{quote}  

\begin{quote}
	\textit{``I  use longer passwords on keyboard, but shorter on touchscreens''} (R757)
\end{quote}  

\begin{quote}
	\textit{``On devices without a user-friendly entry interface, I focus on easily entered passwords or avoid using them altogether.''} (R804)
\end{quote}  

The most common practice (n=35) was simplifying the password by avoiding symbols and numbers.
The next most common practice (n=28) was limiting password length.

While these behaviors make sense from a usability perspective, they are less than ideal from a security perspective.
Additionally, we hypothesize that the number of participants with similar practices is much higher than we identified, as acceptability bias would inhibit participants from mentioning or admitting this behavior.
In fact, a large number of our participants who admitted to this behavior did so sheepishly, admitting that they knew they shouldn't weaken their passwords.

\subsection{Virtual Keyboards}

\begin{table}[htb]
	\centering
	\adjustbox{max width=\columnwidth}{
		\begin{tabular}{ll|r@{\hspace{18pt}} lr|} 	
			&  & \multicolumn{3}{c|}{\shortstack[c]{Count\\(\color{blue}\% Within\color{black};  \% Overall)}} \\
			\midrule
			
			\multirow[c]{2}{*}{\rot{General}} & Any general comment & \multicolumn{3}{c|}{395 (40\%)} \\ \cmidrule{2-5}
			&Negative comments & 315 & (\color{blue} 80\% \color{black};  & 32\%) \\
			&Positive comments & 108 & (\color{blue} 27\% \color{black};  & 11\%) \\
			\midrule
			
			\multirow[c]{6}{*}{\rot{Layout}}
			& Any comment & \multicolumn{3}{c|}{103 (10\%)} \\ \cmidrule{2-5}
			& Familiarity matters & 63 & (\color{blue}61\%; \color{black}  & 6\%) \\
			& Layout switching is hard & 15 & (\color{blue}16\%; \color{black}  & 2\%) \\
			& Finding special characters is hard & 35 & (\color{blue}34\%; \color{black}  & 4\%) \\
			& Capitalization is hard & 7 & (\color{blue}7\%; \color{black}  & 1\%) \\
			\midrule
			
			\multirow[c]{7}{*}{\rot{Arrow Keys}}
			& Any comment & \multicolumn{3}{c|}{222 (22\%)} \\ \cmidrule{2-5}
			& Game: Hard & 107 & (\color{blue}48\%; \color{black}  & 11\%) \\
			& Game: Slow & 38 & (\color{blue}17\%; \color{black}  & 4\%) \\
			& Game: Annoying & 25 & (\color{blue}11\%; \color{black}  & 3\%) \\
			& TV: Hard & 109 & (\color{blue}49\%; \color{black}  & 11\%) \\
			& TV: Slow & 42 & (\color{blue}19\%; \color{black}  & 4\%) \\
			& TV: Annoying & 32 & (\color{blue}14\%; \color{black}  & 3\%) \\
			& Wants cross-device auth & 22 & (\color{blue}10\%; \color{black}  & 2\%) \\
			\midrule

			\multirow[c]{5}{*}{\rot{\shortstack[c]{Mobile \&\\Touchscreen}}} 
			& Any comment & \multicolumn{3}{c|}{187 (19\%)} \\ \cmidrule{2-5}
			& Touchscreen: Easy & 35 & (\color{blue}19\%; \color{black}  & 4\%) \\
			& Touchscreen: Hard & 23 & (\color{blue}12\%; \color{black}  & 2\%) \\
			& Mobile: Easy & 78 & (\color{blue}42\%; \color{black}  & 8\%) \\
			& Mobile: Hard & 65 & (\color{blue}35\%; \color{black}  & 7\%) \\
			& Mobile: Fast & 13 & (\color{blue}7\%; \color{black}  & 1\%) \\
			& Mobile: Slow & 14 & (\color{blue}7\%; \color{black}  & 1\%) \\
			\midrule
		\end{tabular}
	}
	\caption{Themes and codes regarding virtual keyboards. Percentages for codes are reported based on the percentage within the theme and overall.}
	\label{tab:virtual_comments}
\end{table}

Table~\ref{tab:virtual_comments} reports on participants' feelings toward entering passwords and PINs using virtual keyboards.
In total, 395 participants (40\%) commented on this topic.
Overall, the sentiment was negative (86\%; n=315), with participants feeling that typing was much easier using a physical keyboard. %:

%\begin{quote}
%	\textit{``For me, the desktop computer with a keyboard is the easiest device for authentication, because typing on a large keyboard is very comfortable and you can easily type in special characters or symbols that are on the keyboard itself, so that passwords can be long and secure and are easy to type in. On other devices without a physical keyboard, such as a smartphone, it is more difficult to type the characters without making a mistake and it is harder to find the symbols, so the authentication process is slower and produces more errors.} (R424)
%\end{quote}

Participants gave many reasons for not liking virtual keyboards, with many of these comments (n=103) focusing on the layout of the virtual keyboard.
Familiarity with the virtual keyboard's layout was the key concern (n=63), likely explaining why users reported changes with switching between layouts (n=15) or finding special characters (n=35):

\begin{quote}
	\textit{``It's always the easiest for me with the use of [a] keyboard[.] [In the] case of [a] touchscreen it takes more time because of switching keyboards.''} (R11)
\end{quote}
\begin{quote}
	\textit{``[Challenges faced include] finding special characters and switching between capitals and lower case''} (R71)
\end{quote}

Virtual keyboards were especially disliked when participants were required to navigate them using arrow keys (i.e., not a touchscreen), as is the case on most smart TVs and consoles.
222 participants (22\%) indicated that using these devices was difficult, slow, or annoying.
Only a single participant said that it was easy to use these devices.
Further, from section \ref{ch:quan}, 416 participants indicated that they authenticated on a TV and 447 said they had authenticated on a game device before. 
A quarter of these participants (26\%; 24\%, respectively) found authentication on these devices hard.
The following quotes sum up participants' opinions regarding entering passwords using arrow keys:

\begin{quote}
	\textit{``On devices like smart TV where you have to select characters with the TV remote control it is even more complex than on a smartphone, because the system forces you to scroll letter by letter with the remote control until you find the appropriate character, and the process is very slow.''} (R424)
\end{quote}

\begin{quote}
	\textit{``Typing with a TV remote where you have to choose each letter from a grid makes me want to cry.''} (R951)
\end{quote}

To address the challenges associated with entering credentials using arrow keys, several participants (n=22) mentioned wishing they could authenticate on those devices using another device such as their phone (this functionality is supported in some cases):

\begin{quote}
	\textit{``...If you could use your phone or tablet to log into a console instead of using a controller to log in it would make it easier.''} (R460)
\end{quote}
\begin{quote}
	\textit{``Typing a password with a remote/controller can take a lot of time. Some services let you type the password on your phone/PC and then you automatically login on the TV/Console and usually it works well but I wish it was more widespread.''} (R568)
\end{quote}

In contrast to gaming and touchscreen devices, participants' positive remarks around virtual keyboards focused on touchscreens and mobile devices being easy to authenticate on (n=35 and n=78, respectively).
However, even in this case, there were nearly as many comments that said these devices were difficult (n=23 and n=65, respectively).

\subsection{Physical Device Entry}

\begin{table}
	\centering
	\adjustbox{max width=\columnwidth}{
		\begin{tabular}{l|rr@{\hspace{15pt}}l|}
			& \multicolumn{3}{c|}{\shortstack[c]{Count\\(\color{blue}\% Theme\color{black}; \% Overall)}} \\
			\midrule
			Any keyboard comment & \multicolumn{3}{c|}{271 (27\%)} \\ \midrule
			Easy & 221 & (\color{blue}82\%; \color{black}  & 22\%) \\
			Hard & 10 & (\color{blue}4\%; \color{black}  & 1\%) \\
			Fast & 47 & (\color{blue}17\%; \color{black}  & 5\%) \\
			Slow & 0 & (\color{blue}0\%; \color{black}  & 0\%) \\
			Comfortable & 8 & (\color{blue}3\%; \color{black}  & 1\%) \\
			Uncomfortable & 0 & (\color{blue}0\%; \color{black}  & 0\%) \\
			Has less mistakes & 5 & (\color{blue}2\%; \color{black}  & 1\%) \\
			\midrule
		\end{tabular}
	}
	\caption{Codes regarding physical keyboards. Percentages for codes are reported based on the percentage within the theme and overall.}
	\label{tab:physical_keyboard_codes}
\end{table}

In contrast to virtual keyboards being viewed negatively, participants had positive views regarding physical entry (see Table~\ref{tab:physical_keyboard_codes}).
Over two in ten participants (n=221) indicated that physical keyboards found on laptops, desktops, and physical PIN pads were easy to use.
%In \S\ref{ch:quan} we found that mobile devices were the most frequently used device for authentication. 
%This is interesting considering physical devices like laptops and desktops were considered easier to use based on the comments participants left. 
%This indicates that despite their shortcomings in terms of usability, mobile devices have some driving force causing them to be used often. 

\subsection{Authentication Technologies}

\begin{table}
	\centering
	\adjustbox{max width=\textwidth}{
		\adjustbox{max width=\columnwidth}{
			\begin{tabular}{ll|r@{\hspace{18pt}}lr|} 	
				&  & \multicolumn{3}{c|}{\shortstack[c]{Count\\(\color{blue}\% Within\color{black};  \% Overall)}} \\
				\midrule
				
				\multirow[c]{6}{*}{\rot{\shortstack[c]{Password\\managers}}} 
				& Any comment & \multicolumn{3}{c|}{187 (19\%)} \\ \cmidrule{2-5}
				& Uses a password manager & 157 & (\color{blue}84\%; \color{black}  & 16\%) \\
				& Wants a password manager & 22 & (\color{blue}12\%; \color{black}  & 2\%) \\
				& Distrusts password managers & 17 & (\color{blue}9\%; \color{black}  & 2\%) \\
				& Manager is not always available & 67 & (\color{blue}36\%; \color{black}  & 7\%) \\
				& Manager has syncing issues & 6 & (\color{blue}3\%; \color{black}  & 1\%) \\
				\midrule
				
				\multirow[c]{6}{*}{\rot{Biometrics}} 
				& Any comment & \multicolumn{3}{c|}{333 (33\%)} \\ \cmidrule{2-5}
				& Prefers biometrics & 242 & (\color{blue}73\%; \color{black}  & 24\%) \\
				& Dislikes biometrics & 2 & (\color{blue}1\%; \color{black}  & 0\%) \\
				& Distrusts biometrics & 16 & (\color{blue}5\%; \color{black}  & 2\%) \\
				& Biometrics are not always supported & 58 & (\color{blue}17\%; \color{black}  & 6\%) \\
				& Biometrics are innacurate & 117 & (\color{blue}35\%; \color{black}  & 12\%) \\
				\midrule
				
				\multirow[c]{5}{*}{\rot{HSK}} 
				& Any comment& \multicolumn{3}{c|}{71 (7\%)} \\ \cmidrule{2-5}
				& Prefers HSKs & 25 & (\color{blue}35\%; \color{black}  & 3\%) \\
				& Dislikes HSKs & 46 & (\color{blue}65\%; \color{black}  & 5\%) \\
				& Distrusts HSKs & 1 & (\color{blue}1\%; \color{black}  & 0\%) \\
				& HSKs are not always supported & 4 & (\color{blue}6\%; \color{black}  & 0\%) \\
				\midrule
			\end{tabular}
		}
	}
	\caption{Themes and their codes regarding password managers, biometrics, and hardware security tokens (HSKs). Percentages for codes are reported based on the percentage within the theme and overall.}
	\label{tab:password_alternative_codes}
\end{table}

Even though our study focuses on authentication through the use of a password or PIN, nearly half of the participants (n=460) still discussed the roles of password managers, biometrics, and hardware security tokens in their responses.
We report on these results to highlight that users are thinking about these other tools even when they are not mentioned directly, indicating that more research on these tools is needed. 
Their top responses are summarized in Table~\ref{tab:password_alternative_codes}.

\subsubsection{Password Managers}

One in five participants (n=157) mentioned using a password manager.
Several participants (n=22) also mentioned wishing there was a tool that could help them store and generate passwords without being aware that such tools already exist.
The reasons for using a password manager varied but were largely focused on usability (similar to prior work~\cite{pearman2017lets,fagan2017investigation,ray2021why,alodhyani2020password}):

\begin{quote}
	\textit{``ADHD means I have a poor memory. Saving passwords on browser or device helps massively''} (R163)
\end{quote} 

\begin{quote}
	\textit{``I use a password manager which generates complex passwords. These are much faster to enter on a keyboard than a smartphone. This means that it can be time consuming to use a phone and therefore using a desktop computer or even a tablet is preferable.''} (R310)
\end{quote}  

Of particular importance, we note that a third of self-reported password manager users in our sample (n=67) describe frustration related to their password manager not always being available, requiring them to either remember passwords stored there or enter passwords stored in the manager on devices without autofill:

\begin{quote}
	\textit{`I don't remember my passwords because they are all saved on my work computer... when I am trying to use my phone, for example, away from the laptop it is frustrating that I cannot log in''} (R230)
\end{quote}  

\begin{quote}
	\textit{``My phone saves my passwords, however I can't access these passwords via Google Chrome, so I usually make accounts on my phone. If it's a service that I will specifically use on my computer, I will then make an account on my computer...If I use an auto-generated one, I have to refer back to the original device to see what it was. I wouldn't want it to be easier though, as it's safer.''} (R68)
\end{quote}  

Even if they do have the manager on multiple devices, they may run into challenges syncing passwords between devices:

\begin{quote}
	\textit{``Keeping my encrypted password database synced and the versions up-to-date between my mobile and laptop [is a challenge I face].''} (R206)
\end{quote}  

Finally, we note that a small number of participants (n=17) were aware of password managers, but did not trust their security:

\begin{quote}
	\textit{``Having to remember many different combos is difficult, but I don't trust password managers''} (R35)
\end{quote}  

Often, this concern came from a misunderstanding of how password managers work, with participants thinking this caused websites (not managers) to store the plaintext password:

\begin{quote}
	\textit{``When websites are asking if they should save the password its a good system. That allows the user to save the password on websites that the user think are safe.''} (R495)
\end{quote}

\begin{quote}
	\textit{``Remembering all passwords and pins and coming up with a new one that is both strong and easy to remember at the same time is a real modern-day struggle (I don't rely on the ones suggested by Google because I think saving your passwords on a website is extremely unsafe)''} (R614)
\end{quote}  

\subsubsection{Biometrics}

A third of participants (n=333) mentioned biometrics, with the majority of their feedback (n=242) indicating they preferred authentication with biometrics as opposed to entering passwords or PINs.
This is somewhat surprising as this survey did not even mention the topic of biometrics, indicating that this is something participants were passionate about.

The primary reason behind participants' preference for biometrics was their speed, ease of use, and obviating the need to remember a password:
\begin{quote}
	\textit{``I prefer to use fingerprint and not a password because a password can easier be forgotten''} (R91)
\end{quote}  

\begin{quote}
	\textit{``The device I use is easy because I just have to enter my face and it unlocks and also brings up any passwords I may forgotten.''} (R164)
\end{quote}  

Participants liked biometrics so much that five percent of participants (n=58) wished that they could use biometrics for authentication on all of their devices:

\begin{quote}
	\textit{``Using a controller for the Xbox can seem clunkier and harder to use. It would be useful for Xbox to have a fingerprint identification system. ''} (R446)
\end{quote}  

%\begin{quote}
%	\textit{``If i cant use fingerprint sometimes i won't even bother creating [an] account especially if it have long password requirements''} (R317)
%\end{quote}

However, participants did encounter problems (n=117) when using biometrics.
This included biometrics failing to identify the user (false negative) or recognizing another person as the user (false positive):

\begin{quote}
	\textit{``In some lighting facial identification doesn't work\ldots''} (R93)
\end{quote} 

%\begin{quote}
%	\textit{``Sometimes my phone won't read my fingerprint (eg. when it's cold outside)\ldots''} (R125)
%\end{quote} 

\begin{quote}
	\textit{``[I]n the case of unlocking a digital device with a fingerprint, it can be challenging if, amusingly, you aren't mosturizing your hands enough''} (R125)
\end{quote} 

%\begin{quote}
%	\textit{``If I wear my mask it does not recognize me. It was I’ll be easier if it can recognize with and without mask''} (R164)
%\end{quote}  

Moreover, some participants had concerns with the security of biometrics:

\begin{quote}
	\textit{``Im sure face authentication seems uncomortable and sometimes even dangerous.''} (R608)
\end{quote}  

\begin{quote}
	\textit{``Remembering passwords is hell, but I'm uncomfortable with forms of authentication that bypass passwords such as facial recognition and fingerprint authentication, so I feel kind of stuck.''} (R791)
\end{quote}  

\begin{quote}
	\textit{``The fingerprint is not ideal as it is highly easy for anyone to get access to a phone with your fingerprint. At least with people at home or over night visitors. Example: my son placed my thumb on my phone, while I was asleep, and unlocked my phone to play games on my phone...''} (R988)
\end{quote}

\subsubsection{Hardware Security Tokens}
A smaller number of participants (n=77) commented about using hardware security tokens or one-time passwords.
In contrast to biometrics, the most common comment (n=46) regarding hardware security tokens was that participants disliked needing to use a second factor for authentication:

%\begin{quote}
%	\textit{``Multi factor authentication slows the process down.''} (R271)
%\end{quote}

\begin{quote}
	\textit{``I dislike getting verification codes because I recently changed my number and don't have access to my previous phone number anymore.''} (R766)
\end{quote}

\begin{quote}
	\textit{``I don't want to have multiple manual steps to authenticate. If I need to confirm a login I want the confirmation step to be automatic on my device. When an app reads a code from messaging to confirm. I've actually given up logging in when pressed for time on some apps.''} (R882)
\end{quote}

Still, there are some participants (n=25) that preferred the usability of hardware security tokens:

\begin{quote}
	\textit{``It is easier to authenticate using a smart card''} (R447)
\end{quote}

\begin{quote}
	\textit{``[I] find that having to type in number or letter combinations is a bit bothersome, i'd rather use some sort of contactless way to unlock such as a card''} (R563)
\end{quote}

\subsection{Reflections on Passwords}
When asked about their challenges with authentication, nearly half of the participants (n=432) focused on challenges with using passwords.
As these results largely confirm prior work, as opposed to being new findings, we leave the discussion of these results to Appendix~\ref{appx:passwords}.
%The top comments are summarized in Table~\ref{tab:password_codes}.

%% file: chapters/discussion.tex
\section{Discussion} \label{ch:discussion}

Our results demonstrate that users use a wide range of devices to authenticate.
As such, there is a compelling need for more research into improving authentication on non-desktop devices.
Critically, our results demonstrate that this need for research isn't limited to just passwords, but to all forms of authentication, including 2FA and biometrics.

In the remainder of this section, we discuss takeaways and recommendations based on our results.
%In each subsection, we highlight potential future work. 

\subsection{Device-Aware Password Generation}
\label{sec:password-hygiene}
Perhaps the biggest issue presented in this data is that users are changing their passwords based on the device they are using, resulting in weaker passwords for users. 
As respondents R498 and R592 say, the device they use impacts the composition of their passwords.

\begin{quote}
	\textit{``If I do not store confidential data on a given device, I come up with a simple password.''} (R498)
\end{quote}
\begin{quote}
	\textit{``If it's on a device like xbox, ps or nintendo switch, I tend to use a more simple password or a password where the letters are all closeby''} (R592)
\end{quote}

Taken in light of over half of our participants reporting modifying their passwords depending on the device they will use, this suggests that weakening (i.e., shortening or reducing the complexity of) passwords is a common practice.
This is problematic as attackers could more easily compromise these weaker passwords~\cite{weir2010testing}. 

%To highlight how different devices are at the crux of the issue at hand, we consider that prior work by Wash and Rader~\cite{wash2021prioritizing} found little evidence that users consider ease of use when creating passwords when only considering those accounts created for websites.
%Importantly, this shows that users shift their priorities when device type is added as a variable, and it reinforces the significance of considering entry devices when investigating and designing for password entry.

\paragraph{Future Work}
We believe the most effective method for addressing this problem is to design password generation algorithms that are device-aware---i.e., they take into consideration the device where the user will enter the password.
Potential avenues to explore in device-aware password generation include avoiding layout switching on virtual keybaords~\cite{greene2014can,greene2015tap}, prioritizing lowercase characters~\cite{yang2014text}, or preferring long and simple passwords over short and complex passwords.

As a first step in this direction, it will be critical to gather more information on the input characteristics of the devices discussed in this work.
For example, it will be necessary to measure how long it takes to move from letter to letter on a smart TV or how long it takes to switch the virtual keyboard layout on a gaming device.
This fine-grained data will allow for estimating the usability of potential device-aware generation algorithms, helping filter out underperforming schemes before the need for human testing.
For those schemes that appear promising, laboratory and longitudinal studies can confirm usability and test whether improved entry characteristics will ultimately increase the usage of generated passwords.

\subsection{Improving Password Entry Interfaces}
Our study identified several usability issues that can be addressed at the password entry interface.
For example, users mentioned that without knowing the password composition policy (PCP) for a given website, it was difficult for them to remember their password.

Another common usability issue was when password entry interfaces erased incorrect passwords or provided no way for users to view the incorrect password.
This creates a harsh environment where any mistake requires re-entering the entire password, which is especially problematic in the case of touchscreen and directional pad input modalities, which are already slow and cumbersome to use.

\paragraph{Future Work}
An easy step in the right direction is to show the PCP on the password entry page, improving usability with little to no impact on the security.
% Not sure if I should add this because I don't know if it's really "Future Work"
%Another easy implementation for websites is picking more usable PCPs and allowing users to view passwords in plain text as they enter them~\cite{melicher2016usability}. 
While having these features is not as good as using a password manager, it is still something that participants explicitly asked for, and it would help those users who are not yet using a password manager. 
The password entry issue could be easily addressed by persisting incorrect passwords and allowing users the option to view them and correct any mistakes.
Fixing these issues is important as, based on our results (\S\ref{sec:password-hygiene}), improving the usability of password entry could also lead users to select stronger passwords.

%Many users voluntarily brought up PCPs as being confusing or troublesome to deal with. 
%We suggest websites could find ways to help users deal with PCPs, such as restating them when the user is prompted for their password or conforming more to NIST guidelines and allowing length, rather than complexity, to be the metric for password strength~\cite{nist2020digital}.
%
%We find that users are sometimes unable to see the password they are entering (even lacking the ability to toggle a plain text view). 
%When combined with user password attempts being deleted if they are incorrect, 
%Here, we advocate for increased adoption of the ability for users to toggle a plain text view of their password input. 
%Furthermore, future research should determine whether or not deleting incorrect user password attempts is the right course of action.

\subsection{Poor Mental Models}

We were surprised by the number of participants who reported distrusting password managers, biometrics, or hardware security tokens.
While some of this distrust was rooted in an understanding of the security and usability tradeoffs inherent in these different methods, just as much arose due to poor mental models of these technologies.

For password managers, some participants believed that the websites themselves were storing passwords---i.e., not a separate manager application.
This misconception likely arises due to browser-based password managers that often only display their interfaces within the web page and have ambiguous language~\cite{simmons2021systematization}.
For biometrics, the misunderstandings were largely rooted in a misconception that the user's biometric would be sent to a remote server (which would be insecure).

\paragraph{Future Work}
There is room for better communication around how password managers and biometrics work.
Browser-based managers should make it clear to users that it is the browser, not the website, that stores passwords.
Biometrics should also clearly communicate how the biometric is used to authenticate the user.
%Making these changes could have a positive effect on the adoption of password managers and biometrics.
Most likely, this information should be exposed at contextually appropriate times as the user is using the tool~\cite{ruoti2016private}.
Still, we recognize that identifying the correct way to educate users will be non-trivial, likely necessitating significant research in this area. 

\subsection{Service Avoidance and Abandonment}
In our data, we see that users delay using or altogether abandon services if it is difficult to authenticate.
For example, more than half of the participants indicated waiting to create accounts until they were on their preferred devices.
Furthermore, many users explicitly commented that when the process was too hard or slow, they would give up and avoid using a service altogether:

%\begin{quote}
%	\textit{``according to the device I am using, it is more or less quick and easy to authenticate. when it's complicated it's frustrating and sometimes I just give up''} (R387)
%\end{quote}

\begin{quote}
	\textit{``for example, using the remote control to enter a pin is extremely slow and stressful, that is why I often avoid using a platform from the TV if it needs authentication''} (R714)
\end{quote}

Such behavior should be concerning to service providers, who are always looking to attract and keep users.
Perhaps this could be used as a lever to convince these service providers to finally adopt better authentication practices and support more modern forms of authentication~\cite{bonneau2012quest}.

Here, we also note that for many users, the speed of authentication impacted usability more than any other factor, even at times becoming synonymous with usability. 
As such, speed of entry should be a critical concern when designing more usable password generation schemes.
While it is easy to assume this means shorter passwords, it might also involve considering the input device when deciding what passwords would be quick to enter.

\subsection{Other Items of Future Work}

In addition to device-aware password generation, there are other things that password managers could do to improve the usability of password generation.
First, we advocate for a human-centered approach to password generation settings (as opposed to the current entropy-based approach).
When generating a password, instead of asking the user for a password composition policy, it should instead ask the user to indicate how important the account is to them and on which devices they will use the password.
This would allow the creation of passwords that better reflect user needs, while still maintaining an appropriate level of security.

Second, when displaying generated passwords, managers should clearly differentiate character classes when displaying passwords~\cite{simmons2021systematization}.

Lastly, we think it is worth exploring whether password manager functionality could be brought to other devices, such as consoles and smart TVs.
Importantly, we are not calling for consoles or smart TVs to implement password managers as they are implemented on desktops, as taking this approach has already been problematic on mobile phones~\cite{oesch2022observational,seilerhwang2019analyzing}.
Instead, we think there is room for novel approaches that implement only a portion of the manager's functionality, such as autofill without anything else (including password storage).
Research could also explore how this password manager functionality could interact with full password managers, such as by transmitting credentials for autofill over NFC.

\section{Limitations}
\label{sec:limitations}

Our research focused on participants from the Western world.
We did so as our previous attempts to conduct global surveys have been stymied by difficulties translating surveys and survey responses, as well as low participation rates on crowdfunding platforms from non-western countries.
Still, future research should expand this work to explore how it generalizes to different parts of the world.

Many respondents shared their experiences and feedback regarding using biometrics and multi-factor authentication.
Still, we did not directly ask about these topics, meaning that our findings likely underreport them.
Further, a few respondents mentioned the use of passkeys, but because we were focused on authentication that required user input, we did not directly ask about passkeys. 
Future work could conduct a study similar to ours but focused on biometrics, multi-factor authentication, and passkeys.

When asking about passwords, we did not ask respondents to consider the specific account type they were using, though many participants did mention account types in their answers.
Prior research has studied the connection between users choosing passwords, account types, and password reuse~\cite{grawemeyer2011using,florencio2007large,tam2010psychology,gaw2006password,wash2021prioritizing,notoatmodjo2009passwords,haque2013study,duggan2012rational}, and we did not think that asking those questions in this survey would yield new information related to the topic of this paper. 

%Some participants in our study provided feedback about their displeasure with needing a device password or PIN.
%At first glance, this might suggest that they misunderstood our questions, thinking we were concerned with device passwords and PINs, not using passwords and PINs generally.
%Examining our qualitative data, we do not think this was a significant problem, as most users did not exhibit these types of comments, and even many of those that did also provided generic feedback on authentication, suggesting they understood the nature of the question (even if they wanted to complain about device passwords and PINs).
%However, it might still be a minor confounding factor in our data.

Finally, our results are based on self-reported data from users.
There are likely cases where participants misremember details about their device usage.
Similarly, acceptance bias could lead participants to paint an overly positive picture of their authentication practices.
As such, this research should not be considered definitive and will need further studies to further triangulate and confirm our findings.

%% file: chapters/conclusion.tex
\section{Conclusion} \label{ch:conclusion}

The data gathered in our research has helped shed light on the devices users use to enter passwords and PINs, the relative prevalence of authentication using those devices, and users' perceptions and experiences authenticating using these devices.
This will be valuable information to guide the design of improved mechanisms for cross-device credential entry supported by password managers~\cite{simmons2021systematization} and device-aware password generation.
We also identify and discuss challenges that users face using passwords, password managers, and biometrics.
These findings can be used to improve the design of websites and authentication tools.
In particular, we find that there is a need for password managers and biometrics to better educate users about how they function, as many users likely distrust them due to incomplete or inaccurate mental models.

%% file: chapters/appendix-survey.tex
\section{Survey}
\label{appx:survey}

\subsection{Page 1}
In our research group, we are trying to understand on which devices people need to log in to an account. 
\textbf{The process of logging into an account is referred to as authentication.}
We are studying this topic so that in future research we can help make the process of authenticating more seamless, regardless of the device you are using.

Being in this study is up to you. 
After completing the survey, we cannot remove your responses because we will delete any information linking you to your data. 
There are no risks or direct benefits associated with participation in this study. 
Results from this survey will be published in scientific publications. 
Please do not include your name or other identifying information in your survey responses.

If you have questions or concerns about this study, contact us at [email redacted]. 
For questions or concerns about your rights or to speak with someone other than the research team about the study, please contact: [contact information redacted].

\textbf{Statement of Consent}
By continuing in the survey below, I am confirming that I have read the above information and am agreeing to be in this study. 
I can print or save a copy of this consent information for future reference. 
If I do not want to be in this study, I can close my internet browser.

\subsection{Page 2}
\textbf{On which of the following have you entered a password or a PIN?}

\noindent (\textit{Select all that apply})
\begin{itemize}
	\item \textit{Desktop}
	\item \textit{Laptop}
	\item \textit{Phone}
	\item \textit{Touchscreen tablet}
	\item \textit{Smartwatch}
	\item \textit{Smart speaker (e.g. Amazon Alexa)}
	\item \textit{None of the above}
\end{itemize}

\noindent \textbf{On which of the following have you entered a password or a PIN?}

\noindent (\textit{Select all that apply})
\begin{itemize}
	\item \textit{Nintendo Switch}
	\item \textit{Xbox}
	\item \textit{PlayStation}
	\item \textit{Steam Deck}
	\item \textit{VR}
	\item \textit{Other game console} [text entry]
	\item \textit{None of the above}
\end{itemize}

\noindent \textbf{On which of the following have you entered a password or a PIN?}

\noindent (\textit{Select all that apply})
\begin{itemize}
	\item \textit{TV / smart TV}
	\item \textit{thermostat / smart thermostat}
	\item \textit{lock / smart lock}
	\item \textit{safe / smart safe}
	\item \textit{security alarm}
	\item \textit{None of the above}
\end{itemize}

\noindent \textbf{On which of the following have you entered a password or a PIN?}

\noindent (\textit{Select all that apply})
\begin{itemize}
	\item \textit{kiosk computer or tablet}
	\item \textit{printer}
	\item \textit{physical keypad (such as when entering a building)}
	\item \textit{ATM}
	\item \textit{None of the above}
\end{itemize}

\noindent \textbf{Are there any other devices on which you have entered a password or a PIN? Please enter them below.}

\noindent [text entry]

\subsection{Page 3}
\textbf{On which 3–5 devices do you most frequently enter a password or a PIN? Please enter them in order of frequency, from most frequent to least frequent.}
\begin{itemize}
	\item \textit{Device 1} [text entry]
	\item \textit{Device 2} [text entry]
	\item \textit{Device 3} [text entry]
	\item \textit{Device 4} [text entry]
	\item \textit{Device 5} [text entry]
\end{itemize}

\noindent \textbf{How often do you use the following entry methods to enter a password or a PIN?}

\noindent Daily, A few times a week, A few times a month, A few times year, Never
\begin{itemize}
	\item \textit{Keyboard}
	\item \textit{Mouse}
	\item \textit{Touchscreen}
	\item \textit{Physical PIN pad or dial}
	\item \textit{TV remote}
	\item \textit{Video game controller}
\end{itemize}

\subsection{Page 4}
Strongly agree, Agree, Neither agree nor disagree, Disagree, Strongly agree

\noindent I\textbf{ think there is a difference in how easy it is to enter passwords or PINs depending on what device I am using (for example, entering on an Xbox vs entering on a laptop).}

\noindent Strongly agree, Agree, Neither agree nor disagree, Disagree, Strongly agree

\noindent \textbf{When creating passwords or PINs, I consider the types of devices where I will need to enter that password or PIN.}

\noindent Strongly agree, Agree, Neither agree nor disagree, Disagree, Strongly agree

\noindent  \textbf{If I need to create an account, I wait until I can do it on my preferred device type rather than immediately creating the account on the device I am currently using.}

\subsection{Page 5}
\textbf{Please explain how the type of device you are using to enter a password or PIN impacts your experience.}

\noindent  [text entry]

\noindent  \textbf{What challenges do you face when entering passwords or PINs? What do you wish was easier about the process?}

\noindent  [text entry]

\noindent  \textbf{Is there anything else you want to tell us about entering passwords or PINs that could help us improve your experience?}

\noindent  [text entry]

\subsection{Page 6}
\textbf{How old are you?}
\begin{itemize}
	\item \textit{18-25}
	\item \textit{26-35}
	\item \textit{36-45}
	\item \textit{46-55}
	\item \textit{55+}
	\item \textit{I prefer not to enter}
\end{itemize}

\noindent  What is your sex?
\begin{itemize}
	\item \textit{Male}
	\item \textit{Female}
	\item \textit{I prefer not to enter}
\end{itemize}

\noindent  What is your ethnicity?
\begin{itemize}
	\item \textit{White or Caucasian}
	\item \textit{Black or African American}
	\item \textit{Asian}
	\item \textit{Pacific Islander}
	\item \textit{Mixed race}
	\item \textit{Other (specify)} [text entry]
	\item \textit{I prefer not to enter}
\end{itemize}

\noindent What is the highest level of school you have completed or the highest degree you have received?
\begin{itemize}
	\item \textit{Less than high school degree}
	\item \textit{High school graduate (high school diploma or equivalent including GED)}
	\item \textit{Some college but no degree}
	\item \textit{Associate's degree in college (2-year)}
	\item \textit{Bachelor's degree in college (4-year)}
	\item \textit{Master's degree}
	\item \textit{Professional degree (JD, MD)}
	\item \textit{Doctoral degree}
	\item \textit{I prefer not to answer}
\end{itemize}

%% file: chapters/appendix-demographics.tex
\section{Demographics}
\label{appx:demographics}

\begin{figure}
	\centering
	\includegraphics[width=7cm]{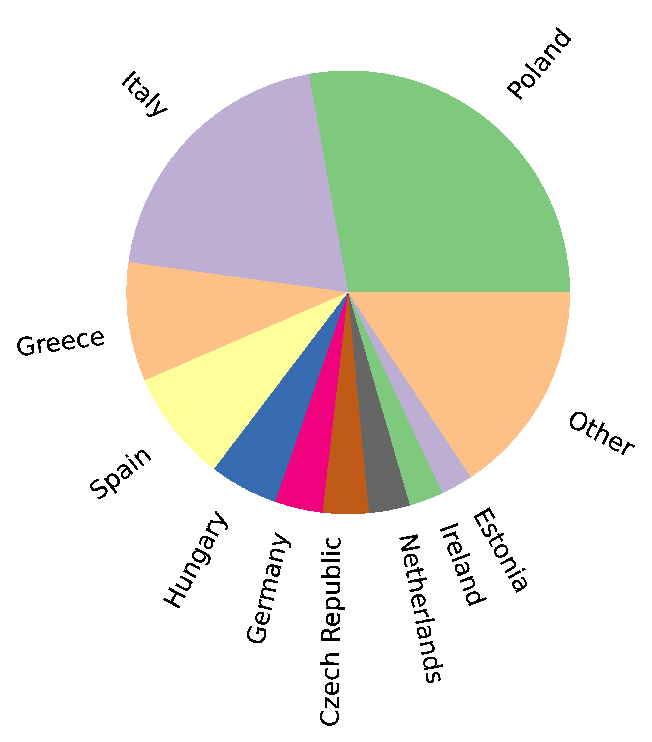}
	\caption{Most common European participants nationality}\label{fig:europe-demographics}
\end{figure}

Table~\ref{tab:demographics} gives overall demographics and Figure~\ref{fig:europe-demographics} shows the breakdown of countries in the European population.

\begin{table*}
	\centering
	\adjustbox{max width=\textwidth}{
		\begin{tabular}{ll|rr|rr|rr|rr|}

			\multicolumn{2}{l|}{} & \multicolumn{2}{c|}{Overall} & \multicolumn{2}{c|}{Europe} & \multicolumn{2}{c|}{USA} & \multicolumn{2}{c|}{UK} \\
			\cmidrule{3-10}
			
			& Participants & 999 & (100\%) & 601 & (60\%) & 299 & (30\%) & 99 & (10\%) \\ \midrule
			
			\multirow[c]{3}{*}{\rot{Gender}} & Male & 495 & (50\%) & 300 & (50\%) & 149 & (50\%) & 46 & (46\%) \\
			& Female & 486 & (49\%) & 292 & (49\%) & 144 & (48\%) & 50 & (51\%) \\
			& I prefer not to answer & 18 & (2\%) & 9 & (1\%) & 6 & (2\%) & 3 & (3\%) \\ \midrule
			
			\multirow[c]{6}{*}{\rot{Age}} & 18-25 & 373 & (37\%) & 282 & (47\%) & 70 & (23\%) & 21 & (21\%) \\
			& 26-35 & 343 & (34\%) & 205 & (34\%) & 103 & (34\%) & 35 & (35\%) \\
			& 36-45 & 161 & (16\%) & 83 & (14\%) & 67 & (22\%) & 11 & (11\%) \\
			& 46-55 & 72 & (7\%) & 22 & (4\%) & 34 & (11\%) & 16 & (16\%) \\
			& 55+ & 47 & (5\%) & 9 & (1\%) & 24 & (8\%) & 14 & (14\%) \\
			& I prefer not to answer & 3 & (0\%) & 0 & (0\%) & 1 & (0\%) & 2 & (2\%) \\ \midrule
			
			\multirow[c]{9}{*}{\rot{Education}} & Less than high school degree & 19 & (2\%) & 12 & (2\%) & 3 & (1\%) & 4 & (4\%) \\
			& High school graduate & 184 & (18\%) & 130 & (22\%) & 41 & (14\%) & 13 & (13\%) \\
			& Some college but no degree & 191 & (19\%) & 94 & (16\%) & 73 & (24\%) & 24 & (24\%) \\
			& Associate's degree in college & 55 & (6\%) & 30 & (5\%) & 20 & (7\%) & 5 & (5\%) \\
			& Bachelor's degree in college & 337 & (34\%) & 192 & (32\%) & 109 & (36\%) & 36 & (36\%) \\
			& Master's degree & 172 & (17\%) & 117 & (19\%) & 41 & (14\%) & 14 & (14\%) \\
			& Doctoral degree & 15 & (2\%) & 11 & (2\%) & 3 & (1\%) & 1 & (1\%) \\
			& Professional degree (JD, MD) & 18 & (2\%) & 9 & (1\%) & 8 & (3\%) & 1 & (1\%) \\
			& I prefer not to answer & 8 & (1\%) & 6 & (1\%) & 1 & (0\%) & 1 & (1\%) \\
			
			\midrule
		\end{tabular}
	}
	\caption{Demographics for the participants taking the study, less those that were removed for quality reasons}
	\label{tab:demographics}
\end{table*}

%% file: chapters/appendix-statistics
\section{Statistics}

We used a series of pairwise $\chi^2$ tests to investigate whether the difference in qualitative feedback between device (see~\S\ref{sec:usability}) was statistically significant.
The results of this analysis are given in Table~\ref{tab:chi-squared}.

\label{appx:chi}
\begin{table*}
	\centering
	\adjustbox{max width=\textwidth}{
		\begin{tabular}{c|c|c|c|c|c|}
			& Physical devices & Game devices & TV remotes & Touchscreen device & Mobile devices \\ \midrule
			Physical devices&\cellcolor{gray}&\cellcolor{gray}&\cellcolor{gray}& \cellcolor{gray}&\cellcolor{gray} \\ \midrule
			Game devices & $\chi^2(7,8)=415.7,p<0.001$ & \cellcolor{gray} & \cellcolor{gray} & \cellcolor{gray} & \cellcolor{gray} \\ \midrule
			TV remotes & $\chi^2(7,8)=432.8,p<0.001$ & $\mathbf{\chi^2(5,6)=3.2382,p=0.66}$ & \cellcolor{gray} & \cellcolor{gray} & \cellcolor{gray}  \\ \midrule
			Touchscreen device & $\chi^2(6,7)=85.79,p<0.001$ & $\chi^2(6,7)=125.7,p<0.001$ & $\chi^2(6,7)=136.5,p<0.001$ & \cellcolor{gray} &\cellcolor{gray} \\ \midrule
			Mobile devices & $\chi^2(6,7)=130.0,p<0.001$ & $\chi^2(6,7)=136.8,p<0.001$ & $\chi^2(6,7)=148.1,p<0.001$ & $\mathbf{\chi^2(5,6)=2.9852,p=0.70}$ &  \cellcolor{gray}  \\ \midrule
		\end{tabular}
	}
	\caption{Table representing the pair-wise $\chi^2$ of the different device categories. $\alpha=0.005$ after apply a Bonferroni correction.}
	\label{tab:chi-squared}
\end{table*}

%% file: chapters/appendix-passwords.tex
\section{Results Relating to Passwords}
\label{appx:passwords}

\paragraph{Password Creation and Memory}

\begin{table}
	\centering
	\adjustbox{max width=\columnwidth}{
		\begin{tabular}{ll|rr@{\hspace{15pt}}l|}
			& & \multicolumn{3}{c|}{\shortstack[c]{Count\\(\color{blue}\% Theme\color{black}; \% Overall)}} \\
			\midrule
			
			\multirow[c]{5}{*}{\rot{Lifecycle}} 
			& Any comment & \multicolumn{3}{c|}{320 (32\%)} \\ \cmidrule{2-5}
			& Remembering passwords is hard & 250 & (\color{blue}78\%; \color{black}  & 25\%) \\
			& Creating passwords is hard & 24 & (\color{blue}8\%; \color{black}  & 2\%) \\
			& Creating unique passwords is hard & 50 & (\color{blue}16\%; \color{black}  & 5\%) \\
			& PCP requirements are troublesome & 76 & (\color{blue}24\%; \color{black}  & 8\%) \\
			\midrule
			
			\multirow[c]{8}{*}{\rot{Usage}} 
			& Any comment & \multicolumn{3}{c|}{143 (14\%)} \\ \cmidrule{2-5}
			& Entry interface malfunction & 59 & (\color{blue}41\%; \color{black}  & 6\%) \\
			& Non-visible password entry & 27 & (\color{blue}19\%; \color{black}  & 3\%) \\
			& Password cleared on mistake & 11 & (\color{blue}8\%; \color{black}  & 1\%) \\
			& Authentication frequency is high & 12 & (\color{blue}8\%; \color{black}  & 1\%) \\
			& Forced password resets & 24 & (\color{blue}17\%; \color{black}  & 2\%) \\
			& Password recovery is hard & 16 & (\color{blue}11\%; \color{black}  & 2\%) \\
			& Account lockout is annoying & 6 & (\color{blue}4\%; \color{black}  & 1\%) \\
			\midrule
		\end{tabular}
	}
	\caption{Themes and codes regarding passwords. Percentages for codes are reported based on the percentage within the theme and overall.}
	\label{tab:password_codes}
\end{table}

A third of participants (n=320) commented about creating and remembering passwords.
A quarter of all participants (n=250) indicated that remembering passwords was their biggest challenge when authenticating:

\begin{quote}
	\textit{``You always end up with more passwords than you want, and you forget them.''} (R50)
\end{quote} 
\begin{quote}
	\textit{``People who created this system need to understand that the average person does not have the mental bandwidth to remember dozens of individual passwords for each site, let alone change them every 90 days or whatever. The whole system is reaching ``peak password'' and I think the whole concept needs to go back to the drawing board.''} (R951)
\end{quote} 

This memory issue led participants to complain that creating new passwords was hard (n=24), particularly regarding creating unique passwords (n=50):

\begin{quote}
	\textit{``I often forget passwords! So I end up using similar ones which of course isn't great for security!''} (R165)
\end{quote}  

Participants (n=76) also complained that password creation and memory are made harder due to password composition policies (PCPs) required by some websites.
Because of these requirements, participants may not be able to use the passwords they want to use.
While this might seem entirely good at first glance (preventing password reuse), it can also prevent users from using password generators as well~\cite{gautam2022improving}.
Participants also complained that because they needed to tailor passwords to the PCP, they struggled to remember these passwords later:

\begin{quote}
	\textit{``Remembering the password as different places require different qualifications for a password, e.g. some require a special character and some don't''} (R277)
\end{quote}  
\begin{quote}
	\textit{``Sometimes it's hard to remember my password, especially when i need to use special characters that i don't use for my other passwords, maybe saying that the password needed a special character so I remember that i needed to add one''} (R426)
\end{quote}  

This frustration was especially pronounced when participants didn't understand why the website needed such strong security:

\begin{quote}
	\textit{``Some [services require] long passwords on platforms [t]hat doesn't need bank level security''} (R317)
\end{quote}  

\paragraph{Password Usage}

One in seven participants (n=143) also mentioned challenges when using passwords.
The most common (n=59) challenge was that authentication hardware (e.g., a PIN pad) or software was often buggy or slow, and sometimes would randomly fail---i.e., \textit{``bugging out and not letting me access what I need''} (R259).

Other complaints centered around the difficulty of correcting mistakes in entered passwords.
This could occur because the entered password was not displayed (e.g., asterisks shown instead of the password) (n=27).
Alternatively, a wrong password would be cleared instead of allowing participants to correct the mistake, necessitating the password to be retyped repeatedly:

\begin{quote}
	\textit{``Sometimes the device won't have the option to let you see the password you entered. I often mistype, so I like to check if I entered it correctly. Also, when I'm not sure if I have the right password in mind and the login fails, it's good to see if it was a simple mistype or if the password itself is incorrect. So I wish more devices had the option to reveal the entered password.''} (R97)
\end{quote} 

\begin{quote}
	\textit{`Sometimes it requires a lot of concentration and focus, since one wrong movement of my finger can make me redo the whole authentication process.''} (R202)
\end{quote} 

Other complaints included that authentication is required too frequently (n=12), that forced password resets are annoying (n=24), or that password recovery (n=16) is too difficult to find or execute:

\begin{quote}
	\textit{``The challenge is that you have many accounts to remember and also in some situations (e-banking i.e.) you have to change password every 3-6 months without using any of the last 10 passwords. That makes the process frustrating.''} (R444)
\end{quote} 
\begin{quote}
	\textit{``It should always there be a way to recover the password or PIN, otherwise someone could be locked out of his/her device/account.''} (R290)
\end{quote}